\documentclass[10pt,letterpaper]{article}
\usepackage{opex3}
\usepackage{subfigure}
\usepackage{amsmath}
\usepackage{graphicx}
\usepackage{epstopdf}
\usepackage{cite}

\newcommand{\asympt}{^{\textrm{as}}}

\begin{document}

\title{Range Dependence of an Optical Pulse Position Modulation Link in the Presence of Background Noise}

\author{Wojciech Zwoli\'{n}ski,$^1$ Marcin Jarzyna,$^{2,*}$ and Konrad~Banaszek$^{2}$}

\address{$^1$Faculty of Physics, University of Warsaw, Pasteura 5, PL-02-093 Warsaw, Poland\\
$^2$Centre of New Technologies, University of Warsaw, Banacha 2c, PL-02-097 Warsaw, Poland}

\email{\textsuperscript{*}m.jarzyna@cent.uw.edu.pl} 



\begin{abstract}
We analyze the information efficiency of a deep-space optical communication link with background noise employing the pulse position modulation (PPM) format and a direct-detection receiver based on Geiger-mode photon counting. The efficiency, quantified using Shannon mutual information, is optimized with respect to the PPM order under the constraint of a given average signal power in simple and complete decoding scenarios. We show that the use of complete decoding, which retrieves information from all combinations of detector photocounts occurring within one PPM frame, allows one to achieve information efficiency scaling as the inverse of the square of the distance, i.e.\ proportional to the received signal power. This represents a qualitative enhancement compared to simple decoding, which treats multiple photocounts within a single PPM frame as erasures and leads to inverse-quartic scaling with the distance. We provide easily computable formulas for the link performance in the limit of diminishing signal power.
\end{abstract}

\ocis{(060.4510) Optical communications; (270.5565) Quantum communications.} 



\section{Introduction}
\label{sec:intro}




Optical domain offers numerous benefits for deep-space communication compared to the radio frequency range \cite{Williams2007}. The primary advantage is access to a much wider bandwidth. Furthermore, the use of laser sources greatly reduces the loss rate due to diffraction of the beam propagating through space, thus allowing for improved targeting of the emitted signal power. Other technical reasons, such as prospectively lesser in size and weight onboard transmitter modules  and the absence of regulatory issues inherent to the use of the radio spectrum additionally make optical communication the technology of choice for future space missions.
This motivates a careful study of the performance limits of optical communication links in the photon-starved regime typical for deep-space scenarios.

\begin{figure}[t]
\includegraphics[width=\linewidth]{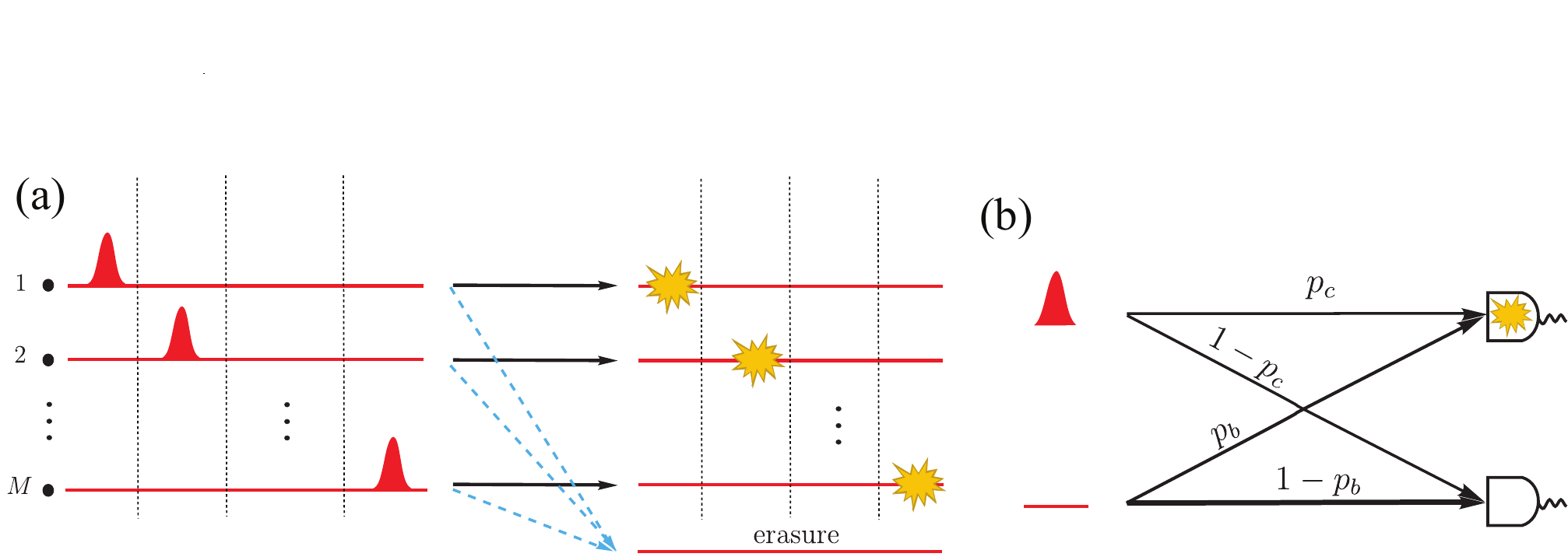}
\caption{(a) The PPM format uses $M$ equiprobable symbols defined by the location of a light pulse in a frame of otherwise empty bins. In the noise-free scenario the input symbol is either identified unambiguously by the timing of the detector click, or erased. (b) In the presence of noise both a light pulse and an empty time bin can generate a detector click with respective probabilities $p_c$ and $p_b$.}\label{Fig:PPMSchematic}
\end{figure}

The standard approach to deep-space optical communication relies on the pulse position modulation (PPM) format shown schematically in Fig.~\ref{Fig:PPMSchematic}(a) which encodes information  in symbols defined by the position of a light pulse within a frame of otherwise empty time bins \cite{Hemmati2005}.
High photon efficiency is achieved by direct detection of the PPM symbols with the help of time-resolved photon counting. In the photon-starved regime some pulses may escape detection, resulting in lower than one probability to generate a click in the bin occupied by a pulse. In the absence of background noise this produces erasures of input PPM symbols which can be efficiently dealt with using standard error correcting codes \cite{Rizzo1997}. Remarkably, it can be shown that with diminishing average signal power the directly-detected PPM format  optimized over the number of time bins within a frame attains the capacity of a narrowband bosonic channel \cite{Giovannetti2014} in the leading order of the power parameter \cite{Guha2011, Waseda2011, Dolinar2011, Kochman, Jarzyna2015, Chung2017}. This is associated with unboundedly growing photon information efficiency as the signal power goes to zero.

The above picture becomes much more nuanced when background noise is taken into account. The common conviction is that in this case the maximum attainable transmission rate scales asymptotically as the inverse of the fourth power of the distance between the transmitter and the receiver \cite{Toyoshima2007, Moision2014}, corresponding to vanishing photon information efficiency. This is quadratically worse compared to conventional coherent communication at radio frequencies, whose information rate exhibits
inverse-square scaling with the distance in the power-limited regime. Such scaling can be viewed as a result of
photon information efficiency attaining a constant value, equal to 1~nat or 2~nats ($1~\text{nat} \approx 1.44~\text{bits}$) respectively for shot-noise limited heterodyne or homodyne detection as implied by the Shannon-Hartley theorem \cite{Proakis}.

The purpose of this paper is to examine carefully using Shannon theory \cite{CoverThomas} the performance limits of an optical communication link based on the PPM format with direct detection in the presence of background noise.
We consider a realistic model of Geiger-mode photon counting detectors which provide only a binary click or no-click outcome in an individual time bin.
We demonstrate theoretically that under the constraint of a given average optical power such a system can in principle achieve inverse-square scaling with the distance while offering at the same time high photon information efficiency. Two ingredients necessary to achieve this regime of operation are identified. The first one is soft decoding strategy which retrieves information from all possible photocount patterns, including multiple clicks within one PPM frame. The second ingredient is the ability to implement the PPM format of an arbitrarily high order, with an increasing number of time bins within one frame. Under the average power constraint this implies unboundedly growing pulse peak power used in the PPM format. Although this requirement may be incompatible with technical limitations of lasers used in onboard transmitters, we point out that the recently presented idea of structured optical receivers \cite{Guha,Rosati2016,Jachura2017} enables one to achieve the same performance with evenly distributed instantaneous power of the transmitted optical signal.

This paper is organized as follows. In Sec.~\ref{Sec:SystemCharacteristics} we review the relevant parameters of a communication link system characteristics. The information rate is calculated in Sec.~\ref{sec:inforate}. Sec.~\ref{Sec:Asymptotic} presents asymptotic analysis of the information rate in the limit of diminishing signal power. These results are used to discuss quantitatively the range dependence of an exemplary PPM link in Sec.~\ref{Sec:RangeDependence}. Finally, Sec.~\ref{sec:discussion} concludes the papers and briefly addresses challenges in implementing the optimized operating mode. 

\section{System characteristics}
\label{Sec:SystemCharacteristics}
The elementary parameters characterizing the transmitter are the emitted signal optical power $P_t$ and the bandwidth $B$, which defines the duration of an individual time bin as $1/B$. Consequently, the average emitted photon number per time bin is $P_t/(Bhf_c)$, where $h$ is Planck's constant and $f_c$ is the signal carrier frequency. Propagation losses and non-unit efficiency $\eta_\text{det}$ of the detector reduce this figure in a linear manner, which yields the average detected signal photon number $n_a$ per time bin given by
\begin{equation}
n_a = \eta_\text{tot}  \frac{P_t}{B h f_c}, \qquad
\eta_\text{tot} = \eta_\text{det} \left(\frac{\pi  f_c D_t D_r}{4 c r}\right)^2.
\label{Eq:n_adef}
\end{equation}
In the above expression $D_t$ and $D_r$ are respectively the diameters of the transmitter and the receiver antennas, $r$ is the distance covered by the optical link, and $c$ denotes the speed of light. Conveniently, $n_a$ is a dimensionless parameter characterizing the strength of the detected signal. Eq.~(\ref{Eq:n_adef}) describes diffractive losses which make $n_a$ scale as $r^{-2}$ with the range when all other parameters of the link are fixed. We will be interested in the photon-starved regime, when $n_a \ll 1$, arising for large distances.

The $M$-ary PPM format uses $M$ equiprobable symbols corresponding to the location of a single light pulse in a frame of $M$ otherwise empty bins shown in Fig.~\ref{Fig:PPMSchematic}(a). In order to satisfy the average power constraint, the mean photon number in the pulse needs to be equal to $n_s = Mn_a$. Without background noise, direct detection identifies unambiguously the input symbol through the timing information, unless the photon counting detector does not fire at all over the duration of the PPM frame. According to the standard theory of photodetection \cite{Kelley1964} the probability of such an erasure event is $\exp(-n_s)$. From the information theoretic viewpoint the communication scheme is described by an $M$-ary erasure channel with the probability of faithful transmission equal to $1-\exp(-n_s)$.

In the presence of background noise, photocounts may occur also in empty time bins.
The noise model considered in this work is based on an assumption that stray light and dark counts generate background whose strength is equivalent to $n_b$ photons per time bin and that background counts are statistically independent from each other as well as uncorrelated with the incoming signal. Furthermore, we will take a realistic model for photon counting which discriminates only between the presence or absence of clicks in a given time bin, which applies e.g.\ to avalanche photodiodes operated in the Geiger mode. Thus the detector generates a click in an empty time bin and that occupied by a light pulse with respective probabilities
\begin{equation}
\label{Eq:pbpcdef}
p_b = 1- \exp(-n_b), \qquad p_c = 1- \exp(-n_s - n_b).
\end{equation}
The conditional probabilities for a single time bin are depicted schematically in Fig.~\ref{Fig:PPMSchematic}(b). Other noise models, such as single-mode thermal fluctuations \cite{Jarzyna2018}, can by analyzed by replacing Eq.~(\ref{Eq:pbpcdef}) with suitable alternative expressions and following steps described below.

\section{Information rate}
\label{sec:inforate}
The most elementary decoding strategy for a noisy link is to interpret as erasures all events when clicks have occurred in multiple time bins within one PPM frame. Such simple decoding would either recover the input PPM symbol, although with a certain error probability induced by background counts, or yield an erasure event. A more general soft decoding scenario would be to retrieve information also from sequences containing multiple clicks in individual PPM frames. The maximum attainable transmission rate is given by the Shannon mutual information evaluated for the statistics of output events taken into consideration. The probability of obtaining a specific sequence of clicks in exactly $k$ time bins is equal to one of two expressions
\begin{equation}\label{eq:prob_n}
p_c(k)=p_c p_b^{k-1}(1-p_b)^{M-k},\qquad p_e(k)=(1-p_c)p_b^{k}(1-p_b)^{M-k-1}
\end{equation}
depending respectively on whether the signal pulse was located in one of those time bins or not.
For a given value of $k$ there are respectively ${M-1 \choose k-1}$ and ${M-1 \choose k}$ combinations of clicks in time bins for each sequence type. The marginal probability of observing a given sequence of $k$ clicks for any of the $M$ equiprobable input PPM symbols is given by a weighted sum
\begin{equation}
p(k) = \frac{k}{M} p_c(k) + \left( 1- \frac{k}{M} \right) p_e(k)
\end{equation}
with the total number of sequences containing $k$ clicks among $M$ bins equal to ${M \choose k}$.

\begin{figure}
\centerline{\includegraphics[width=\linewidth]{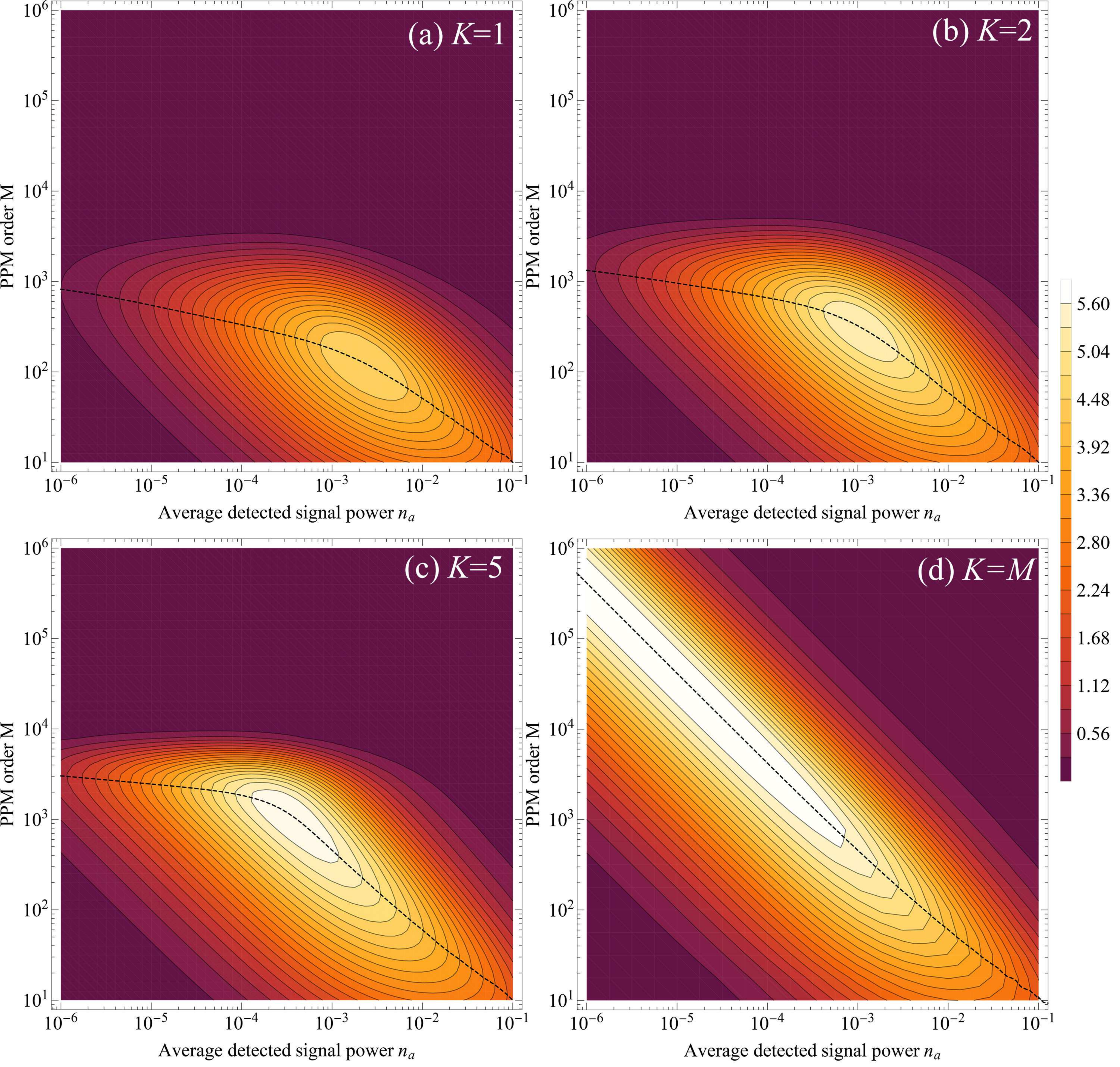}}
\caption{Photon information efficiency $I^{(K)}/n_a$ as a function of the average detected signal power $n_a$ and the PPM order $M$ for a fixed background noise strength $n_b = 10^{-3}$. Results are shown for decoding truncated at (a) $K=1$, (b) $K=2$, (c) $K=5$ as well as (d) complete decoding with $K=M$. The dashed curves indicate optimal PPM order $M^\ast$ as a function of $n_a$.}
\label{Fig:PIEContour}
\end{figure}

In order to take into account general soft decoding strategies, we will evaluate mutual information per time bin $I^{(K)}$ for a scenario when information is retrieved from sequences containing up to $K$ clicks, while other events are interpreted as erasures. The complete expression reads:
\begin{equation}\label{eq:I_K}
I^{(K)}= \frac{1}{M}  \sum_{k=1}^K \left[ {M-1 \choose k-1} p_c(k)\log_2 p_c(k)
+ {M-1 \choose k} p_e(k)\log_2 p_e(k)-  {M\choose k}p(k)\log_2 p(k) \right].
\end{equation}
The first two terms stem from the average conditional entropy of the output when the input symbol is known, whereas the last term is contributed by the entropy of the output variable itself. Simple decoding corresponds to the case $K=1$, whereas for complete decoding $K=M$. In the following discussion it will be convenient to use the photon information efficiency (PIE) given by the ratio $I^{(K)}/n_a$, which specifies the amount of transmitted information per one detected photon.

In Fig.~\ref{Fig:PIEContour} we present contour plots of PIE as a function of the PPM order $M$ and the detected signal power $n_a$ for a fixed background noise power $n_b = 10^{-3}$. Decoding restricted at a fixed level exemplified with $K=1,2,5$ is compared to the complete decoding scenario when $K=M$. The qualitative difference between these two cases is clearly seen. While for restricted decoding PIE tends to zero with $n_a \rightarrow 0$, complete decoding enables one to attain a non-zero asymptotic value of PIE with an appropriate choice of the PPM order. This advantage of complete decoding is associated with a divergent asymptotic behaviour of the optimal PPM order $M^\ast$ with the vanishing signal power, shown  in Fig.~\ref{Fig:PIEContour} with dashed lines.

\begin{figure}
\centerline{\includegraphics[width=\linewidth]{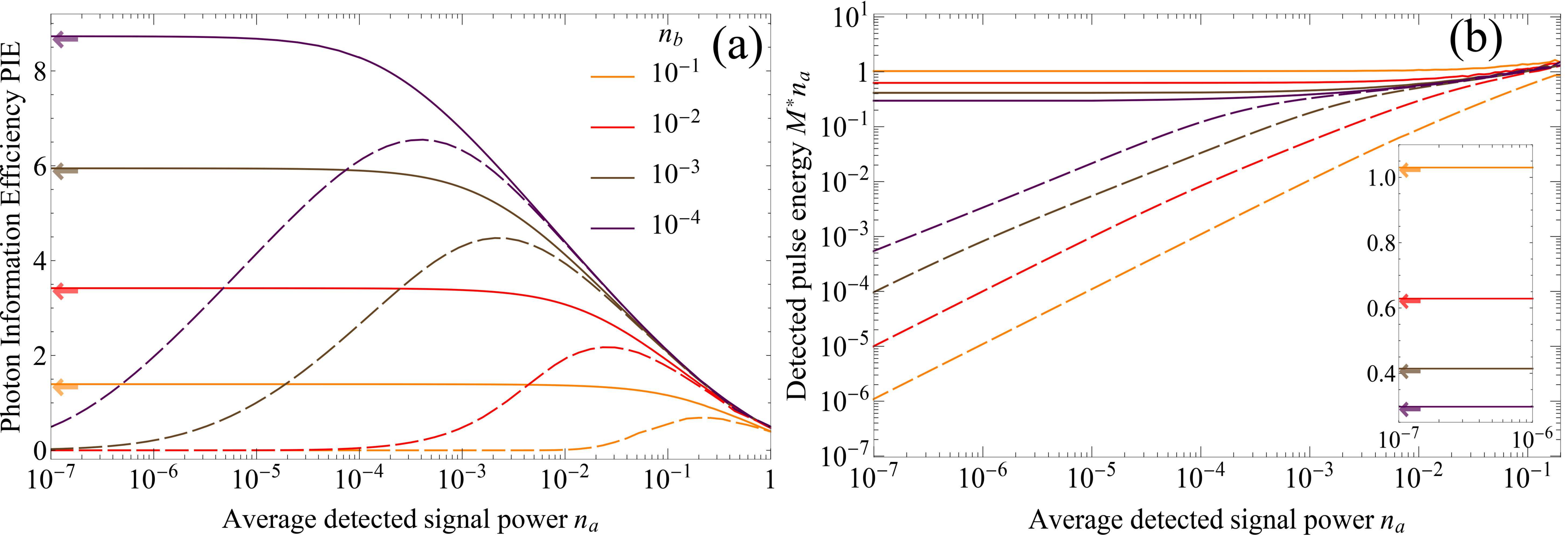}}
\caption{(a) Photon information efficiency optimized over the PPM order for complete decoding $\text{PIE}^\ast = \max_{M}(I^{(M)}/n_a)$ (solid lines) and simple decoding $\max_{M}(I^{(1)}/n_a)$ (dashed lines), shown as a function of the average detected signal power $n_a$ for several values of the background noise strength $n_b$. (b) The corresponding optimal pulse detected optical energy for complete decoding $n_s^\ast = M^\ast n_a$ (solid lines) and simple decoding (dashed lines).
Arrows in the panel (a) and in the inset of the panel (b) indicate asymptotic values for the complete decoding scenario calculated using Eq.~(\ref{Eq:AsymptoticPIE}).}
\label{Fig:PIEOpt}
\end{figure}

In order to gain further insights into the performance of the complete decoding scenario, in Fig.~\ref{Fig:PIEOpt} we plot the
maximum attainable photon information efficiency $\text{PIE}^{\ast}$ and the corresponding optimal pulse optical energy $n_s^\ast$ given by
\begin{equation}
\text{PIE}^{\ast} = \max_{M}(I^{(M)}/n_a), \qquad n_s^\ast = M^\ast n_a
\end{equation}
as a function of the average detected optical power $n_a$ for several values of the background noise parameter $n_b$. It is seen that in the limit $n_a \rightarrow 0$, both $\text{PIE}^{\ast}$ and $n_s^\ast$ tend to constant values which depend on the background noise. The arrows shown in Fig.~\ref{Fig:PIEOpt}(a) indicate the asymptotic values of the PIE calculated using the method presented in Sec.~\ref{Sec:Asymptotic}. This method provides also the asymptotic values of $n_s^\ast$, indicated with arrows in the inset of Fig.~\ref{Fig:PIEOpt}(b). Consequently, in the asymptotic limit the optimal PPM order $M^\ast = n_s^\ast/n_a$ scales inversely with the detected optimal power $n_a$ as seen in Fig.~\ref{Fig:PIEContour}(d). The results for complete decoding are in stark contrast with the simple decoding strategy shown for comparison in Fig.~\ref{Fig:PIEOpt} with dashed lines. In the latter case, both the photon information efficiency and the optimal pulse energy tend to zero as $n_a \rightarrow 0$.

\section{Asymptotic PIE value}
\label{Sec:Asymptotic}

In this section we present a simple method to calculate the asymptotic values of the maximum photon information efficiency $\text{PIE}^{\ast}$ and the corresponding optimal pulse optical energy $n_s^\ast$ for the complete decoding scenario. The presented results are based on the information theoretic analysis of the channel capacity per unit cost \cite{Verdu1990}.
The starting observation is that the $M$-ary PPM format can be viewed as a constrained version of generalized on-off keying (OOK) with a binary set of elementary symbols, where a light pulse is sent with a probability $1/M$ and an empty bin with the probability $1-1/M$. The constraint has the form of a requirement that every sequence of $M$ consecutive time bins contains exactly one pulse. Because of this constraint, the mutual information $I^{(M)}$ for the completely-decoded PPM link is upper-bounded by the mutual information $I_{\text{OOK}}$ for generalized OOK. The latter can be written as \cite{CoverThomas}
\begin{equation}\label{Eq:IOOK}
I^{(M)} \le I_{\text{OOK}} =  \left( 1 - \frac{1}{M} \right)
D \left( p_b || (1- M^{-1}) p_b + M^{-1} p_c \right) + \frac{1}{M} D \left( p_c || (1- M^{-1}) p_b + M^{-1} p_c \right)
\end{equation}
where $p_b$ and $p_c$ are defined in Eq.~(\ref{Eq:pbpcdef}) and
\begin{equation}
\label{Eq:RelEntdef}
D(p||q)=p\log_2\frac{p}{q}+(1-p)\log_2\frac{1-p}{1-q}
\end{equation}
denotes the relative entropy between binary probability distributions $p,1-p$ and $q,1-q$.
Consequently, for a given $n_a$ the photon information efficiency $\text{PIE}^\ast$ for the PPM format optimized over the order $M$ will be less or equal to $\text{PIE}^\ast_{\text{OOK}} = \max_M(I_{\text{OOK}}/n_a)$, where in the last expression maximization is carried out over $M$ taken as a continuous positive parameter. Next, an upper bound on $\text{PIE}^\ast_{\text{OOK}}$ can be obtained from the information theoretic result on the channel capacity per unit cost \cite{Verdu1990}. For the communication link analyzed here the average cost is measured in terms of the average optical energy per time bin and there is available exactly one input symbol with cost equal to zero, namely the empty time bin. In this setting the capacity per unit cost is monotone non-increasing in $n_a$ and its asymptotic value $\text{PIE}\asympt$ in the limit $n_a \rightarrow 0$ can be obtained from the following single-parameter maximization recipe:
\begin{equation}
\label{Eq:AsymptoticPIE}
 \text{PIE}\asympt = \max_{n_s\geq 0}\frac{D(p_c||p_b)}{n_s},
\end{equation}
where the ratio $D(p_c||p_b)/n_s$ on the right hand side of the above equation should be expressed using Eq.~(\ref{Eq:pbpcdef}) in terms of two parameters: $n_b$, treated as a given constant, and $n_s$ being the optimization variable.

\begin{figure}
\centerline{\includegraphics[width=\linewidth]{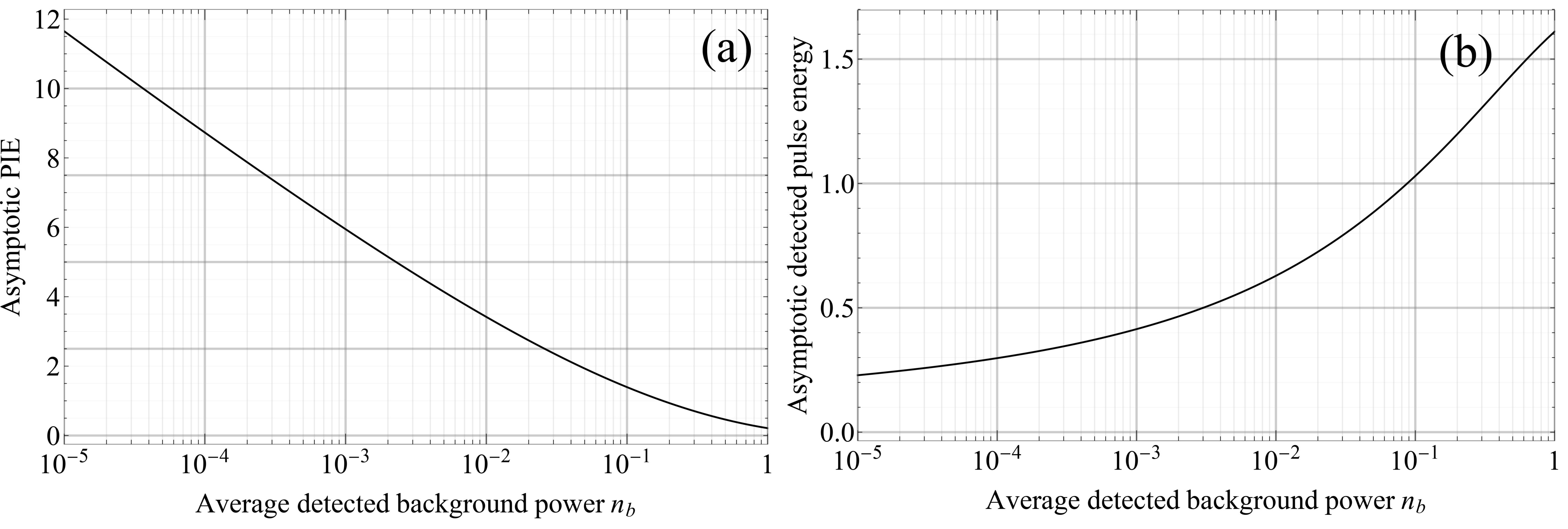}}
\caption{(a) The asymptotic photon information efficiency $\text{PIE}\asympt$ for a completely decoded PPM link in the limit of the vanishing average detected signal power $n_a \rightarrow 0$ as a function of the background noise strength $n_b$. (b) The corresponding optimal detected pulse energy $n_s\asympt$. The depicted values have been obtained using Eq.~(\ref{Eq:AsymptoticPIE}).}
\label{Fig:PIEAsymptotics}
\end{figure}

The reasoning presented above implies that $\text{PIE}\asympt$ defined in Eq.~(\ref{Eq:AsymptoticPIE}) specifies an upper bound on the photon information efficiency for the PPM format,
\begin{equation}
\text{PIE}^{\ast}\le \text{PIE}^\ast_{\text{OOK}} \le \text{PIE}\asympt.
\end{equation}
Furthermore, we show in Appendix that the value $\text{PIE}\asympt$ is actually attained in the limit $n_a \rightarrow 0$ by the photon information efficiency of a PPM link in the complete decoding scenario. In Fig.~\ref{Fig:PIEAsymptotics} we plot $\text{PIE}\asympt$ as a function of the background noise strength $n_b$. The figure also shows the optimal $n_s\asympt$ maximizing the right hand side of Eq.~(\ref{Eq:AsymptoticPIE}). Let us recall that $n_s\asympt$ characterizes the {\em detected} optical energy of the pulse. The values $\text{PIE}\asympt$ and $n_s\asympt$ characterize the attainable long-range performance of a completely-decoded PPM link under a constraint of a fixed average detected signal power. This performance will be discussed in more detail in Sec.~\ref{Sec:RangeDependence}.

For completeness, we will close this section by analyzing the asymptotic limit $n_a \rightarrow 0$ of the simple decoding scenario.
Fig.~\ref{Fig:PIEOpt}(b) indicates that in the case of simple decoding the optimal pulse energy $n_s$ tends to zero with the diminishing average power $n_a$. This observation motivates expanding the mutual information $I^{(1)}$ into a power series in $n_s$. The leading order term has quadratic dependence on $n_s$,
\begin{equation}
I^{(1)} \approx \log_2 e \frac{(M-1)e^{-(M-1)n_b}}{2 M^2 (1-e^{-n_b})} n_s^2.
\end{equation}
Inserting $n_s = M n_a$ yields the photon information efficiency for a given PPM order approximately equal to
$I^{(1)}/n_a \approx (\log_2 e) (M-1)e^{-(M-1)n_b} n_a / [2(1-e^{-n_b})]$. With $M$ treated as a continuous parameter, the right hand side of the above formula is maximized by $M = 1 + n_b^{-1}$, which gives:
\begin{equation}
\text{PIE}^{(1)} \approx \frac{\log_2 e}{2 e n_b (1-e^{-n_b})} n_a.
\label{Eq:PIE(1)}
\end{equation}
Thus the optimized photon information efficiency scales linearly with the average detected optical power $n_a$, which is clearly visible in Fig.~\ref{Fig:PIEOpt}(a) form the behavior of dashed curves. The corresponding optimal detected pulse energy, given approximately by
\begin{equation}
n_s^{(1)} \approx (1+n_b^{-1})n_a
\label{Eq:ns(1)}
\end{equation}
also exhibits asymptotic linear scaling in $n_a$, clearly seen in Fig.~\ref{Fig:PIEOpt}(b).

\section{Range dependence}
\label{Sec:RangeDependence}

Based on results derived in the preceding sections, the maximum information rate $R^\ast$ of a PPM link characterized by the bandwidth $B$ can be written as:
\begin{equation}
R^\ast = B \cdot n_a  \cdot \text{PIE}^\ast(n_a, n_b),
\label{Eq:Rstar}
\end{equation}
where $n_a$ is the average detected signal power per time bin. For clarity, we explicitly stated here the dependence of the photon information efficiency $\text{PIE}^\ast$ optimized with respect to the PPM order on $n_a$ as well as the background noise strength $n_b$. The actual dependence has been depicted in Fig.~\ref{Fig:PIEOpt}(a). Assuming for simplicity that the optical pulse employed in the PPM format has a rectangular shape filling the entire time bin with duration equal to $B^{-1}$, the peak power requirement to attain the optimal performance reads:
\begin{equation}
P_{\text{peak}}^\ast = B \cdot \eta_{\text{tot}}^{-1} \cdot  h f_c \cdot n_s^\ast(n_a, n_b),
\label{Eq:Ppeakstar}
\end{equation}
where $\eta_{\text{tot}}$ is the overall link transmission efficiency including detection losses defined in Eq.~(\ref{Eq:n_adef}), $hf_{\text{c}}$ is the energy of a single photon at the carrier frequency, and $n_s^\ast(n_a, n_b)$ is the optimal detected pulse energy shown in Fig.~\ref{Fig:PIEOpt}(b).

\begin{figure}
\centerline{\includegraphics[width=0.85\linewidth]{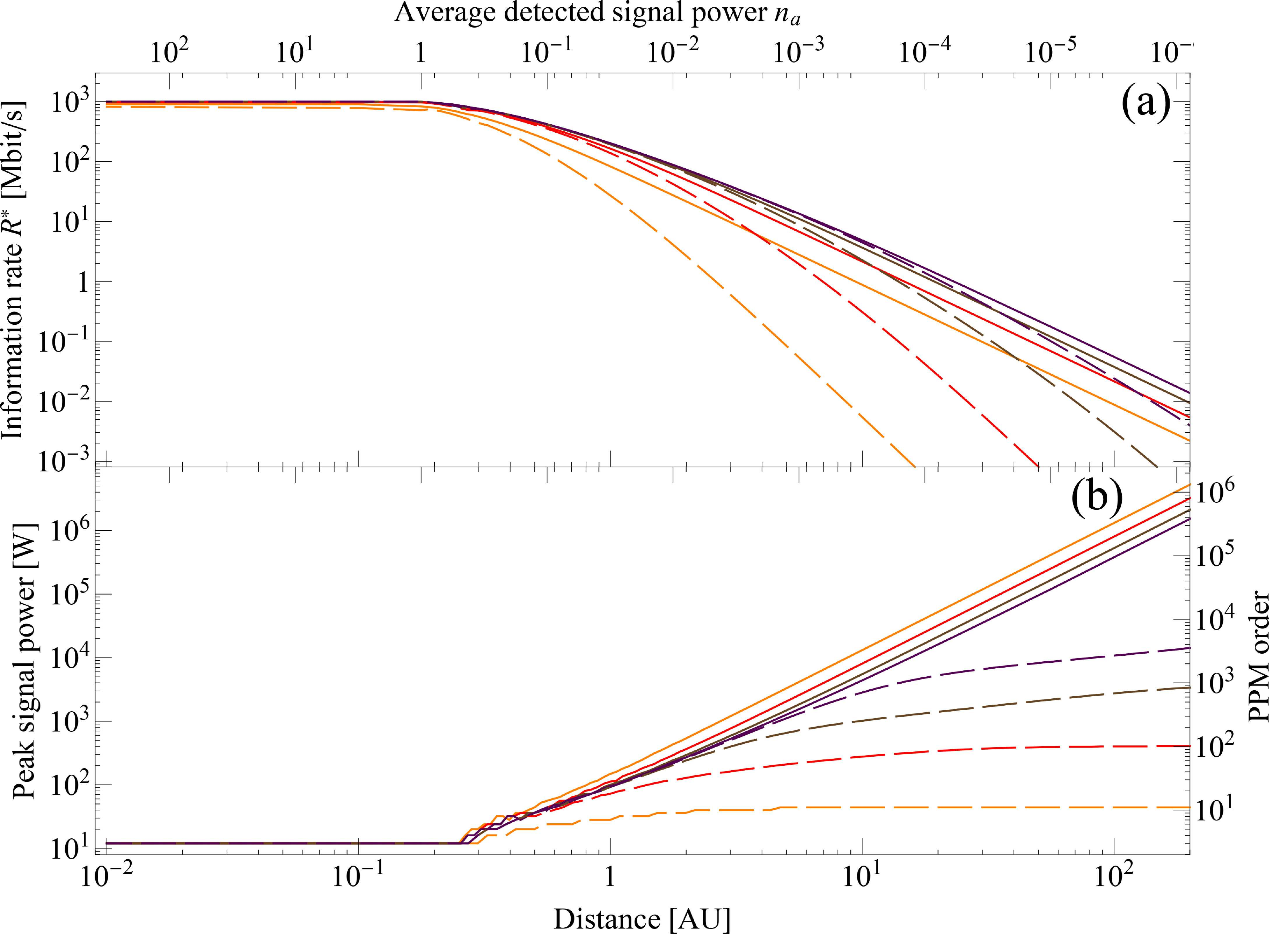}}
\caption{Maximum information rate of PPM optical communication link (a), optimal PPM order (right scale) and peak input signal power (left scale) (b) as a function of distance $r$ for complete (solid lines) and simple (dashed lines) decoding for various level of noise and setup parameters $f_c=2\cdot 10^5$ GHz, $P=4$ W, $B=2$ GHz, detector efficiency $\eta_{\textrm{det}}=0.025$ and transmitter and receiver antenna diameter $D_t=0.22$ m and $D_r=11.8$ m respectively. $1\,$AU$\approx 1.5\cdot 10^8$km.}
\label{Fig:RangeDependence}
\end{figure}

The link range $r$ enters Eqs.~(\ref{Eq:Rstar}) and (\ref{Eq:Ppeakstar}) through the parameters $\eta_{\text{tot}}$ and $n_a$, both defined in Eq.~(\ref{Eq:n_adef}). As a numerical example, we have taken the transmitter optical power $P_t = 4~\text{W}$, the link bandwidth $B = 2~\text{GHz}$, the carrier frequency $f_c = 2 \cdot 10^{5}~\text{GHz}$, and the transmitter and the receiver antenna diameters respectively $D_t = 0.22~\text{m}$ and $D_r = 11.8~\text{m}$. For this set of parameters, the attainable information rate $R^\ast$, the optimal PPM order $M^\ast$, and the required peak power $P_{\text{peak}}^\ast$ are shown in Fig.~\ref{Fig:RangeDependence} with solid lines as a function of the link range expressed in astronomical units (AU) for several values of the background noise parameter $n_b$. For short ranges, below approximately $0.2~\text{AU}$ the performance of the link is limited by the available bandwidth. In this regime the information rate can be characterized by the expression for the noise-free model, given by $B\cdot M^{-1} \log_2 M$. The optimal performance is achieved by the ternary PPM format with $M=3$, which gives a slightly higher value of mutual information $M^{-1} \log_2 M \approx 0.528$~bit/bin compared to either binary ($M=2$) or quaternary ($M=4$) formats for which mutual information is $0.5$~bit per time bin.

For ranges beyond several AU, the information rate $R^\ast$ in the complete decoding scenario shown in Fig.~\ref{Fig:RangeDependence}(a) exhibits a very favorable dependence with the distance $r$ following $r^{-2}$ scaling analogous to that of the detected signal power. This behavior stems from the fact that for  diminishing signal power the photon information efficiency $\text{PIE}^\ast(n_a, n_b)$ in Eq.~(\ref{Eq:Rstar}) approaches the constant value $\text{PIE}\asympt$ depending only on the noise strength $n_b$, and the link range $r$ enters the right hand side of Eq.~(\ref{Eq:Rstar}) only through $n_a$. Achieving this performance requires the implementation of extremely high PPM orders, as seen in Fig.~\ref{Fig:RangeDependence}(b). The required PPM order is given by $M^\ast = n_s^\ast(n_a, n_b)/n_a$ and for large distances it scales as $r^2$, as in this regime the optimal detected pulse energy $n_s^\ast(n_a, n_b)$ becomes only a function of the noise strength $n_b$. The same scaling is exhibited by the peak power $P_{\text{peak}}^\ast$ evaluated according to Eq.~(\ref{Eq:Ppeakstar}) and shown in Fig.~\ref{Fig:RangeDependence}(b).

The above results are juxtaposed  in Fig.~\ref{Fig:RangeDependence} with the simple decoding scenario depicted with dashed lines. Most importantly, for long ranges the attainable information rate exhibits disadvantageous $r^{-4}$ scaling with the distance. This behavior can be easily understood by inserting in Eqs.~(\ref{Eq:Rstar}) and (\ref{Eq:Ppeakstar})  in lieu of $\text{PIE}^\ast$ and $n_s^\ast$ the asymptotic expressions for $\text{PIE}^{(1)}$ and $n_s^{(1)}$ derived respectively in Eqs.~(\ref{Eq:PIE(1)}) and (\ref{Eq:ns(1)}). Because $\text{PIE}^{(1)}$ is linear in $n_a$, the information rate exhibits quadratic scaling with $n_a$ implying $r^{-4}$ dependence on the distance. On the other hand, the optimal PPM order tends to a constant value for large distances and so does the peak power. The numerical difference between complete and simple decoding is significant, for example at $r=10~\text{AU}$ and the background noise strength $n_b=10^{-1}$ complete decoding allows one to increase the information rate by a factor of thousands.

\section{Discussion}
\label{sec:discussion}
We have analyzed the range dependence of a noisy optical communication link employing the PPM format and a Geiger-mode photon counting detector, which produces a binary click or no-click outcome in each elementary time bin. Under a fixed average signal power constraint, the attainable system performance, quantified using Shannon information, dramatically depends on the adopted decoding strategy. In the complete decoding scenario, when information is retrieved from all detection events including sequences containing multiple clicks within one PPM frame, it is in principle possible to achieve $r^{-2}$ scaling of the information rate with the distance, i.e.\ the rate becomes directly proportional to the detected signal power. However, the optimal operating regime requires a careful adjustment of the PPM order to the system characteristics, growing as $r^{2}$ with the covered distance.

The necessity to generate the signal in the form of infrequent strong pulses may lower the overall electrical-to-optical power conversion efficiency of the transmitter module, which is essential for downlink space communication. This issue can be resolved by the use of recently proposed structured optical receivers \cite{Guha,Rosati2016,Jachura2017}. The basic idea is to generate the optical signal with evenly distributed instantaneous optical power in the form of carefully designed phase or phase-and-polarization patterns which enable one to concentrate temporally the signal energy after transmission using optical interference. Such schemes with quasi-cw optical signals can achieve the efficiency of the PPM format at the expense of a more complicated construction and operation of the receiver. However, these are secondary considerations for downlink transmission which is the main bottleneck in deep-space communication.

\section*{Acknowledgments}
We acknowledge insightful discussions with C. Heese, M. Jachura, and M. Srinivasan.

\section*{Funding}
This work is part of the project ``Quantum Optical Communication Systems'' carried out within the TEAM programme of the Foundation for Polish Science co-financed by the European Union
under the European Regional Development Fund.

\appendix

\section*{Appendix}

In order to derive the lower bound on the photon information efficiency $\text{PIE}^\ast= \max_{M}(I^{(M)}/n_a)$ in the complete decoding scenario it will be helpful to resort to an alternative form of the mutual information $I^{(M)}$. Let us denote the count sequence within one PPM frame as $y_1y_2\ldots y_M$, where $y_j = 0$ denotes no detector click in the $j$th time bin, while $y_j=1$ labels a click in that bin. Further, let $p(y|0)$ with
\begin{equation}
p(0|0) = 1-p_b, \quad p(1|0) = p_b
\end{equation}
be the conditional no-click and click probabilities for an empty time bin and $p(y|1)$ with
\begin{equation}
p(0|1) = 1-p_c, \quad p(1|1) = p_c
\end{equation}
denote analogous probabilities for a bin containing the light pulse. Because all PPM symbols are equiprobable and differ only by the position of the light pulse, mutual information for the complete set of count sequences can be written as
\begin{equation}
I^{(M)} = - \frac{1}{M} \sum_{y_1,y_2,\ldots, y_M =0,1} p(y_1|1) p(y_2|0) \ldots p(y_M|0)
\log_2 \left[ \frac{1}{M} \left( 1 + \frac{p(y_1|0)}{p(y_1|1)} \sum_{j=2}^{M} \frac{p(y_j|1)}{p(y_j|0)} \right) \right].
\end{equation}
The convexity of the function $-\log_2 (ax+b)$ in the argument $x$ for positive $a$ and $b$ allows one to apply Jensen's inequality to individual summations over parameters $y_2, y_3, \ldots, y_M$ which yields
\begin{equation}
I^{(M)} \ge -\frac{1}{M} \sum_{y=0,1} p(y|1)\log_2 \left[ \frac{1}{M} \left( 1 + \frac{p(y|0)}{p(y|1)} (M-1) \right) \right].
\end{equation}
The right hand side can be written in terms of the relative entropy defined in Eq.~(\ref{Eq:RelEntdef}) as
\begin{equation}
I^{(M)} \ge \frac{1}{M} D \left( p_c || (1- M^{-1}) p_b + M^{-1} p_c \right)
\end{equation}
where we have returned to notation used in the main text. Let us note that the right hand side is identical with the second term on the right hand side of Eq.~(\ref{Eq:IOOK}). Consequently,
$
\text{PIE}^\ast \ge  D \left( p_c || (1- M^{-1}) p_b + M^{-1} p_c \right)/({M n_a})
$
for any $M$. Let us now insert $M=n_s\asympt/n_a$, where $n_s\asympt$ is the value maximizing the right hand side of Eq.~(\ref{Eq:AsymptoticPIE}). This yields:
\begin{equation}
\text{PIE}^\ast \ge \frac{1}{n_s\asympt} D \left( p_c\asympt || (1-n_a/n_s\asympt) p_b +  n_a p_c\asympt / n_s\asympt \right).
\end{equation}
where $p_c\asympt$ is evaluated for $n_s\asympt$.
Because relative entropy is continuous, in the limit $n_a \rightarrow 0$ one has
$\text{PIE}^\ast \ge D(p_c\asympt || p_b)/ n_s\asympt$.
Together with Eq.~(\ref{Eq:AsymptoticPIE}), the above inequality implies that $\text{PIE}\asympt$ is the asymptotic value of photon information efficiency also for the optimized PPM format.

\end{document}